# Spontaneous parametric down-conversion

Christophe Couteau[1]

[1]University of Technology of Troyes
Light, Nanotechnologies & Nanomaterials (L2n)
ICD CNRS, 12 rue Marie Curie, 10000 Troyes, France.

Spontaneous Parametric Down-Conversion (SPDC), also known as parametric fluorescence, parametric noise, parametric scattering and all various combinations of the abbreviation SPDC, is a non-linear optical process where a photon spontaneously splits into two other photons of lower energies. One would think that this article is about particle physics and yet it is not, as this process can occur fairly easily on a day to day basis in an optics laboratory. Nowadays, SPDC is at the heart of many quantum optics experiments for applications in quantum cryptography, quantum simulation, quantum metrology but also for testing fundamentals laws of physics in quantum mechanics. In this article, we will focus on the physics of this process and highlight few important properties of SPDC. There will be two parts: a first theoretical one showing the particular quantum nature of SPDC and the second part, more experimental and in particular focusing on applications of parametric down-conversion. This is clearly a non-exhaustive article about parametric down-conversion as there is a tremendous literature on the subject, but it gives the necessary first elements needed for a novice student or researcher to work on SPDC sources of light.

Even though SPDC has been predicted in the 1960s [1] and first demonstrated experimentally by Burham and Weinberg (from NASA!) in 1970 [2], this process was really used for practical purposes only in the late 80s and early 90s. Since then, this process is at the heart of many quantum optics experiments for applications in quantum cryptography [3], quantum computing [4], quantum metrology [5] but also simply for testing fundamentals laws of physics in quantum mechanics [6]. This review article is clearly not exhaustive as a massive literature exists on the subject and cannot be summarized in 20 pages only. Rather, this article intends to provide the necessary basic formalism and knowledge there is to know if one wants to work with SPDC and what it is good for. We will not go on to explain how to build your own SPDC source as plenty of literature already exists on this matter [7,8]. In that context, we will first develop some basic non-linear optics derived from Maxwell equations with a focus on the exact reverse process of SPDC, namely second-harmonic generation (SHG). Then we will go on to introduce some quantum optics in a nutshell before going into the theoretical derivation of SPDC. In the second part, we will develop what SPDC is used for with the notion of correlated photons [9], indistinguishable photons [10], entangled photons [11] but also how it can be used to engineer a single photon source [12]. To finish, we will give some examples of how parametric fluorescence was used for the demonstration of quantum teleportation [13], quantum cryptography [14], CNOT gates [15], quantum candela [16] or for quantum lithography [17]. We will finish by giving some perspectives of engineering new sources of SDPC in the hope of increasing the efficiency of this physical effect.

### I- Theoretical derivation of SPDC: classical case

Without going into the details of non-linear optics, let us refresh our mind with some concepts [18]. The idea is that when an electromagnetic (em) wave interacts with a medium full of charges, there is a dipolar-type of interaction between the dipoles (charges) in the medium and the incoming em field at the frequency ω. The classical harmonic oscillator model holds which is the model of an electron attached to an atom-core behaving like a spring and thus like an harmonic oscillator interacting with the incoming wave. We then obtain the standard linear complex refractive index of the medium, which would depend on ω. But if one starts to increase significantly the energy of the incoming field then this harmonic oscillator will start to be anharmonic *i.e.* not respond linearly with the incident electromagnetic wave. One can simulate this behaviour by the following equation of motion for an electron in the medium under such em wave by:

$$\frac{d^2x}{dt^2} + 2\gamma \frac{dx}{dt} + \omega_0^2 x + ax^2 = F = -\frac{e}{m}E(t) \qquad (1)$$

where the first term on the left hand side is the motion acceleration contribution, the second term is a damping term that tends to slow down oscillations mainly due to non-radiative contribution and the last two terms on the left side of the equality cause the restoring force of the electrons towards its initial position. The anharmonicity is contained in the last term ($a.x^2$) which we can neglect in linear optics. If ignored, then one can see that the dependence of the motion equation is linear with the position x. This is no longer true at high em power, then the electromagnetic force given by the right-hand side of equation (1) will induce non-linearities (we note that higher terms than $a.x^2$ can also exist). Now if one takes two different incoming frequencies $\omega_1$ and $\omega_2$, one can show that this equation of motion, thanks to the last quadratic term, will lead to the generation of waves at $\omega_1$ and $\omega_2$ of course, but also at $\omega_1 + \omega_2$, $\omega_1 - \omega_2$, $2\omega_1$, $2\omega_2$ and 0 (so-called optical rectification) leading to 6 frequency components [18]. We can solve this equation by assuming that the non-linear term is a perturbation case of the linear one in order to end up with these 6 frequencies. A particularly interesting case occurs when $\omega_1=\omega_2=\omega$ as this will lead to the so-called second-harmonic generation at 2ω, so useful in every optics laboratory. Higher terms can also exist leading to other non-linear phenomena with more interacting waves. This classical approach is very useful to understand the physics of the interaction but is rapidly complex when using the perturbation theory. There is another equivalent way of treating the problem by starting with Maxwell's equations with the displacement term $\vec{D}$ and the magnetic field strength $\vec{H}$ in a medium which is not the vacuum. We suppose no external charges, no currents and we suppose a non-magnetic medium, thus we have the standards relations $\vec{D} = \varepsilon_0 \vec{E} + \vec{P}$ for the displacement, where $\vec{P}$ is the polarisation vector of the medium, and $\vec{B} = \mu_0 \vec{H}$ for the magnetic field. The polarisation vector $\vec{P}$ accounts for the usual charge displacement in a medium due to an applied external em field. By playing around with Maxwell's equations, one can show the following known wave equation:

$$\vec{\nabla}^2 \vec{E} - \frac{1}{c^2}\frac{\partial^2 \vec{E}}{\partial t^2} = \frac{1}{\varepsilon_0 c^2}\frac{\partial^2 \vec{P}}{\partial t^2} \qquad (2)$$

with the polarisation $\vec{P} = \vec{P}^L + \vec{P}^{NL}$ containing both the linear (L) and the non-linear (NL) terms. SPDC (and SHG for that matter) is a three-wave interaction only and thus we will only consider the first non-linear term for the polarisation given by:

$$P_i^{NL} = \varepsilon_0 \sum_j \sum_k \chi_{ijk}^{(2)} . E_j E_k \qquad (3)$$

with $\{i, j, k\} = \{x, y, z\}$, meaning that the second order term $\chi_{ijk}^{(2)}$ is actually a tensor and is called the second order nonlinear susceptibility of the medium. Figure 1-a) represents the Feynman's diagram of the SPDC process where the photon γ₃ at energy $\hbar\omega_3$ splits into two twin photons (γ₂ and γ₁) at energies $\hbar\omega_2$ and $\hbar\omega_1$. SHG would be the exact time-reverse process. Even though very odd looking at a first glance, SPDC does follow energy conservation and momentum conservation rules so that:

$$\hbar\omega_3 = \hbar\omega_1 + \hbar\omega_2 \qquad (4)$$

$$\vec{k}_3 = \vec{k}_2 + \vec{k}_1 \text{ and } \Delta\vec{k} = \vec{k}_3 - \vec{k}_2 - \vec{k}_1 = \vec{0} \qquad (5)$$

These two equations are called the *phase-matching conditions* and will be discussed in details later on as they are central for the mechanism to happen. These conditions are represented in Figure 1-b (energy conservation) and Figure 1-c (momentum conservation). Now one can solve the wave equation from Eq. (2) with the usual assumptions where we consider a lossless medium, collimated and monochromatic input beams, continuous excitation, normal incidence and we will neglect the double refraction. We will also make a less-reasonable assumption which is to be in the non-depleted regime meaning that the incoming beam can give rise to SHG (reverse diagram than the one presented in Figure 1-a) but without significant energy loss for the excitation beam or so (the so-called pump beam, with 2 photons at frequency ω coming from the same pump laser in this case). This assumption can be quite wrong as the incoming beam can be more or less completely depleted for SHG if one is careful with phase-matching conditions.

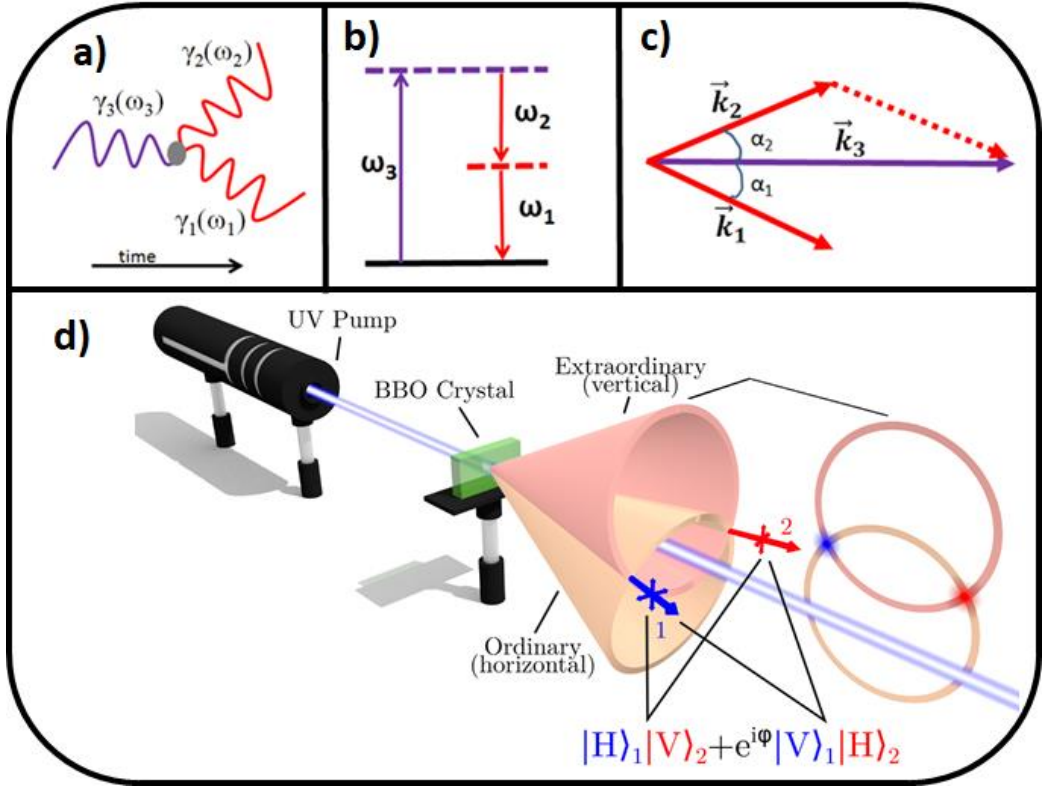

**Figure 1: a)** Feynman's diagram of SPDC where one photon at energy $\hbar\omega_3$ splits into two twin photons at energies $\hbar\omega_2$ and $\hbar\omega_1$ and **b)** representation of energy conservation of SPDC, **c)** is the representation of momentum conservation of SPDC and **d)** is a schematic of SPDC where a UV pump laser generates red SPDC photons coming out of the crystal in a cone shape.

For SPDC on the other hand, we are (unfortunately) completely in the regime of no-pump depletion as parametric down-conversion is an extremely inefficient process. Under such assumptions, one can solve the equation (2) by looking for a solution of the form:

$$E_3(z,t) = A_3 e^{i(k_3 z - \omega_3 t)} + c.c. \qquad (6)$$

'c.c.' being the complex conjugate term and $k_3 = n_3 \cdot \omega_3/c$ and $n_3^2 = \varepsilon^L(\omega_3)$ where *L* stands for the linear part only. We thus look for a solution for the polarisation of the form:

$$P_3(z,t) = \wp_3 \cdot e^{i(k_3 z - \omega_3 t)} + c.c. \qquad (7)$$

with

$$\wp_3 = 4\varepsilon_0 d_{eff} E_1 E_2 \qquad (8)$$

$E_1$ and $E_2$ being the amplitudes of the em fields 1 and 2. $d_{eff} = 1/2 \chi_{eff}$ is a complicated but known parameter depending on the strength of the non-linearity of the material and on geometrical factors all of which are properties of the material (as an example, for BBO type II, we have $d_{eff} = (d_{11} \sin(3\phi) + d_{22}\cos(3\phi))\cos(2\theta)$). One can then find the following equation from equations (2-3) and (6-8):

$$\frac{d^2 A_3}{dz^2} + 2ik_3 \frac{dA_3}{dz} = -\frac{4 d_{eff} \omega_3}{c^2} A_1 A_2 e^{i(k_1 + k_2 - k_3)z} \qquad (9)$$

This last equation can be further reduced by assuming the slow varying amplitude approximation and thus neglect the first term compared to the second term (also a standard approximation in non-linear optics) on the left-hand side of this equation. It is then straightforward to find the solution of such equation under the non-depleted pump approximation by integration between 0 and L (length of the crystal) and considering that $I_i = 2n_i\varepsilon_0 c |A_i|^2$ is the optical intensity of beam $i$:

$$I_3 = \frac{8d_{eff}^2 \omega_3^2 I_1 I_2}{n_1 n_2 n_3 \varepsilon_0 c^3} L^2 \left(\frac{\sin \Delta k.L/2}{\Delta k.L/2}\right)^2 = I_3^{max}. sinc^2(\Delta k.L/2) \qquad (10)$$

with the so-called phase-matching parameter $\Delta k = k_1 + k_2 - k_3$ (we use scalar notations for easier use). One can see from equation (10) that ideally, one would want to have $\Delta k \sim 0$ in order to have a quadratic increase (first order Taylor expansion of the *sinc* function) of the SHG intensity as a function of the length of the crystal. Otherwise, the *sinc* function ensures a periodic energy exchange between the em pump field (the incoming beam) and the created waves. This phase-matching condition is crucial for an efficient non-linear effect and accounts for the wave vector conservation of the process. In order to get the phase-matching condition right, one needs to have:

$$k_1 + k_2 = k_3 = \frac{n_3 \omega_3}{c} = \frac{n_1 \omega_1}{c} + \frac{n_2 \omega_2}{c} \qquad (11)$$

which is an impossible condition to be satisfied with most materials as $n_1(\omega_1) < n_2(\omega_2) < n_3(\omega_3)$ for $\omega_1 < \omega_2 < \omega_3$, *i.e.* for 'normal' dispersion. This is where birefringent crystals are used as they possess two (uniaxial) or three (biaxial) different refractive indices along different symmetry axes which can satisfy the phase-matching condition. In the particular case of second-harmonic generation, it is fairly easy to have $n_e(\omega_1) = n_o(\omega_3 = 2\omega_1)$ with *o* and *e* the so-called ordinary and extraordinary axes for uniaxial non-linear crystals.

SPDC is the reverse process of sum-frequency generation (SFG which is non-degenerate SPDC, *i.e.* different wavelengths for the two output photons) or SHG (degenerate SPDC, *i.e.* same wavelengths for the two output photons) which would correspond to the process called parametric amplification in classical non-linear optics. This process occurs when a strong pump beam at $\omega_3$ is combined with a weak signal beam at $\omega_1$ to give rise to an amplified signal at $\omega_1$ but also to a second signal called the idler beam at $\omega_2$. Using the same type of equation than the one derived in (9), we have this time:

$$\frac{dA_1}{dz} = i \frac{\omega_1}{n_1 c} \chi_{eff}^{(2)} A_3 A_2^* e^{i\Delta k.z} \qquad (12)$$

$$\frac{dA_2}{dz} = i \frac{\omega_2}{n_2 c} \chi_{eff}^{(2)} A_3 A_1^* e^{i\Delta k.z} \qquad (13)$$

These equations leading to the so-called Manley-Rowe relations can be solved supposing that $\Delta k = 0$ and lead to:

$$A_2(z) = i \sqrt{\frac{\omega_2 n_1}{\omega_1 n_2}} \frac{A_3}{|A_3|} A_1^*(0) \sinh(\alpha z) \qquad (14)$$

$$A_1(z) = A_1(0) \cosh(\alpha z) \qquad (15)$$

$$\alpha = \frac{\chi_{eff}^{(2)} |A_3|}{c} \sqrt{\frac{\omega_2 \omega_1}{n_2 n_1}} \qquad (16)$$

From these solutions of the parametric amplification process, we can clearly see that if there is no initial incident power at frequency $\omega_1$, *i.e.* if $A_1(0) = 0$, then the signal and the idler beams would not exist and SPDC would not be possible. In other words, classical non-linear optics "does not allow" spontaneous parametric down-conversion to exist. Clearly, we see the limitations of the classical treatment of non-linear optics in order to derive SPDC. For that, a quantum description of the phenomena is needed [19,20].

## II- Theoretical derivation of SPDC: quantum case

In order to do so, let us recall some basic quantum optics properties. Quantum optics is the realm of electromagnetism and Maxwell's equations when one deals with single photons. Indeed, a process such as SPDC tells you that spontaneously, a photon pair can suddenly 'appear' somewhere in the universe ex-nihilo. Once again, energy and momentum conservation hold but nevertheless, a photon in a particular optical mode can spontaneously be created. In order to account for such phenomena (and many others in quantum optics), one needs to quantise the electromagnetic field. This article is certainly not about that and one can find excellent textbooks on the matter [21]. We will just give a quick overview of what is required to know. The general principle is to start with Maxwell's equations and use a spatial Fourier expansion of the em field in order to solve these equations. It is also wiser to work with the vector potential $\vec{A}$ rather than directly with the electric field $\vec{E}$. If one works in the so-called Coulomb gauge where $\vec{\nabla}.\vec{A} = 0$ (which tells you that photons are transverse!) then one can end up with an equation that looks very much like the one of an harmonic oscillator with the derivation of conjugate canonical variables just like in the case of a quantised harmonic oscillator. It is then straightforward to apply the rules of a quantum harmonic oscillator but this time for the vector potential of the electromagnetic field. By doing so, one can show the following relation:

$$\vec{A}(\vec{r}) = \sum_l \vec{\epsilon}_l \frac{\mathcal{E}_l}{\omega_l} \left( e^{i\vec{k}_l.\vec{r}} \hat{a}_l + e^{-i\vec{k}_l.\vec{r}} \hat{a}_l^\dagger \right) = \vec{A}^{(+)}(\vec{r}) + \vec{A}^{(-)}(\vec{r}) \qquad (17)$$

with the sum over the $l$ optical modes accessible in this universe (quite a few then) and the two quantum operators, $\hat{a}$ being the photon annihilation operator (accounting for the disappearance of a photon) and $\hat{a}^\dagger$ the photon creation operator (accounting for the appearance of a photon). By the same token, $\vec{A}$ is also an operator. $\vec{\epsilon}_l$ is the vector direction of the em field associated to these photons and $\mathcal{E}_l$ is the amplitude of the associated classical field with energy $\hbar\omega_l$ equals to that of a single photon in the $l$ mode. $\hat{a}$ and $\hat{a}^\dagger$ are the equivalent creation and annihilation operators of a quantum of energy for an harmonic oscillator, allowing to go up and down in the energy ladder of the oscillator, which in the case of quantum optics, is called a photon. We have then the following relations:

$$\hat{a}_l^\dagger |vac\rangle = |1_l\rangle \qquad (18)$$

$$\hat{a}_l |1_l\rangle = |vac\rangle \qquad (19)$$

where a single photon in mode $l$ is created from the vacuum state $|vac\rangle$ and a single photon in mode $l$ "disappears" for instance due to absorption from a single dipole. One can show that the quantised free radiation Hamiltonian would look like:

$$\hat{H}_R = \sum_l \hbar\omega_l \left( \hat{N}_l + \frac{1}{2} \right) = \sum_l \hbar\omega_l \left( \hat{a}_l^\dagger \hat{a}_l + \frac{1}{2} \right) \qquad (20)$$

Where $\hat{N}_l$ is the photon number operator giving rise to the number $N_l$ of photons in the electromagnetic mode $l$. Within this context, quantum optics allows and explains perfectly the possibility of spontaneously creating a photon of a certain energy in a given optical mode. Thus a photon of high energy can split into two (or more but less likely) photons of lower energies that were not previously around when the interaction with the non-linear medium takes place. In order to describe the phenomena, we can infer an effective interaction Hamiltonian for SPDC which would look like:

$$\hat{H}_{SPDC} = i\hbar\kappa \left( \hat{a}_1 \hat{a}_2 \hat{a}_3^\dagger e^{i\Delta\vec{k}.\vec{r} - i\Delta\omega.t} + \hat{a}_1^\dagger \hat{a}_2^\dagger \hat{a}_3 e^{-i\Delta\vec{k}.\vec{r} + i\Delta\omega.t} \right)$$

$$= i\hbar\kappa \left( \hat{a}_i \hat{a}_s \hat{a}_p^\dagger e^{i\Delta\vec{k}.\vec{r} - i\Delta\omega.t} + \hat{a}_i^\dagger \hat{a}_s^\dagger \hat{a}_p e^{-i\Delta\vec{k}.\vec{r} + i\Delta\omega.t} \right) \quad (21)$$

with the first term of the Hamiltonian accounting for SFG or SHG (if $\omega_1 = \omega_2$) and the creation of a SHG photon at $\omega_3 = 2\omega_2$ with $\Delta\omega = \omega_3 - \omega_1 - \omega_2$. The second term of the Hamiltonian accounts for SPDC where one photon (the pump beam) 'disappears' to create one photon at $\omega_1$ and one photon at $\omega_2$ in the non-degenerate case where the two frequencies are different. We can clearly see in (21) that SPDC and SHG do compete and do exist at the same time. In most of what follows in this article, we will focus on degenerate SPDC with $\omega_1 = \omega_2$ thus 'giving birth' to twin photons in this case. Note that from now on, we will use the indices $\{s, i, p\}$ rather than $\{1,2,3\}$ when talking about the three-wave mixing phenomenon SPDC. We have $p$ for the incoming pump photon at high energy and the twin photons $i$ for idler and $s$ for signal (note that these names are given more for historical reasons). We must stress that the constant $\kappa$ is given by:

$$\kappa = \frac{2}{3} \frac{d_{eff}}{\varepsilon_0 V} \sqrt{\frac{\omega_p \omega_s \omega_i}{2\varepsilon_0 V}} \quad (22)$$

with the usual meanings for $d_{eff}$, $V$ and the 3 frequencies. Now in a more concrete way we can estimate what happens to an incoming pump beam with $N_p$ number of pump photons impinging on a non-linear crystal by looking at the effect of the SPDC Hamiltonian on the incoming state $|0_s, 0_i, N_p\rangle$ describing $N_p$ pump photons and 0 photon in the signal mode and 0 photon in the idler mode. We use the Schrödinger equation in time and look for the state:

$$|\psi(t)\rangle = e^{\frac{1}{i\hbar}\int_0^t \hat{H}_{SPDC}(t')dt'} |0_s, 0_i, N_p\rangle \quad (23)$$

$\kappa$ is rather small so it means that SPDC is very inefficient and that most of the pump beam is unperturbed by the SPDC process. We can thus make a Taylor expansion of the exponential Hamiltonian up to the first order and find that:

$$|\psi(t)\rangle \approx C_0 |0_s, 0_i, N_p\rangle + C_1 \frac{1}{i\hbar} \int_0^t \hat{H}_{SPDC}(t')dt' |0_s, 0_i, N_p\rangle + \cdots \quad (24)$$

where $C_i$ are the coefficients of the Taylor expansion present for the normalization of the wave function. In the case of 1st order, we only have $C_0$ and $C_1 \approx 1 - C_0$ again for the sake of working with normalised states. As we are looking for perfect phase-matching, including in energy too, then we have $\Delta\omega \approx 0$ thus the integral term in (24) is a Dirac function and we can apply the operators $\hat{a}_i^\dagger, \hat{a}_s^\dagger, \hat{a}_p$ on the incoming state $|0_s, 0_i, N_p\rangle$ and find:

$$|\psi(t)\rangle = C_0|0_s, 0_i, N_p\rangle + \kappa C_1 e^{-i\vec{\Delta k}.\vec{r}}|1_s, 1_i, N_p - 1\rangle \qquad (25)$$

In order to bring out a true parallel with the SHG case (from Equation (10)), we must recall that the measured em intensity $I_{SPDC}$ will be proportional to the square of the wave function from (25). Thus the phase matching condition in wave-vector does appear (like in SPDC). Unlike SHG, SPDC is a linear function with the pump power which is not trivial to see at first when looking at (25). For that, we need to go back to (21) and assume that the pump power is so strong that it can be considered like a classical em field $\hat{a}_p \approx E_p$ and as a result, $I_{SPDC}$ will be proportional to $I_p$ only, unlike the SHG phenomena from (10). For SPDC, one pump photon only splits into two photons but only one photon in each mode so the dependence is clearly linear even though it is a non-linear process. We also recall that $C_0 \gg C_1$. This is actually one of the major issues of SPDC, it is the fact that SPDC is a very inefficient process which hinders many applications in quantum information and quantum cryptography for instance. We also assumed that we only have the first 2 terms in the Taylor expansion of (24) but if $\hat{a}_p \approx E_p$ is strong enough (sometimes necessary if one wants to get enough SPDC effect) then higher order terms appear. This expansion is thus a Poisson distribution that characterizes SPDC and thus not an on-demand supply of photon pairs but rather a probabilistic one. From (25), we see that for a certain probability to have one pair (proportional to $\kappa^2$) we have a non-zero chance to have two pairs (with probability to be $\kappa^4/4$) when one add the second order term etc… In other words, most of the incoming state (the pump beam) is left untouched and only a small portion (10$^{-5}$ to 10$^{-12}$) goes into producing a twin pair of down-converted photons. We also see that the double-pair generation increases if the pump power is increased too much and that is not something one would necessarily want for most experiments using single twin photon pairs. Finally, we must stress that in formula (25), we 'increase' the probability of being in the $|1_s, 1_i, N_p - 1\rangle$ by performing a so-called coincidence measurement between the idler and the signal photon in a quantum optics experiment. Practically, we look at a double-detection event on two detectors (one pointing towards the idler photons and the other one towards the signal photons) in a very short time-scale of 10 ns or so. One can show that accidental events in such a short time window are very low and thus a coincidence event corresponds to the detection of a pair of SPDC photons. This is the practical way of getting rid of the vacuum state, or rather the initial state $|0_s, 0_i, N_p\rangle$, where we collapse the wave function unto the one-and-more-pair-coincidence state, thus getting rid of the contribution from this dominant non-creation of pairs state. Mandel *et al.* also demonstrated that the number of events for parametric down-conversion has to the same as for the pump photons. In other words, one pump photon might split into two SPDC photons but the number of pair-event created has to be at the most the number of incoming photons, assuming every photon is converted and no higher order occur. It is obvious experimentally but not trivial. Finally, unlike the reverse case of SHG which requires a non-zero 'seed' signal $A_1(0)$ in order for any SPDC photons to exist, thanks to the properties of the creation operators, we can start from zero photons in the idler and signal states and create them using the Hamiltonian in (21). To wrap up on the theory of SPDC, we can claim that the fact that only quantum formalism can explain the existence of SPDC photons, is actually one of the many peculiar properties of parametric fluorescence and that what makes it unique for applications in quantum optics as we will learn later on.

### III- Practical case: BBO crystal for SPDC

Practically in the laboratory there are many aspects that should be taken care of in order to generate 'efficient' SPDC in an optical set-up. First of all, the phase-matching condition $\Delta k = 0$ has to be fulfilled as much as possible. As it was mentioned earlier it is quite unlikely to find a natural material which fulfill the condition $n_i(\omega_i) = n_s(\omega_s) < n_p(\omega_p)$ with $\omega_i = \omega_s < \omega_p$ except when using birefringent crystals which may possess two (uniaxial crystals) or three (biaxial crystals) different indices for a given wavelength. Another important point is that we want non-collinear phase-matching leading to different wave vectors directions for the three waves involved thus more easily selected thereafter (see schematic Fig. 1-d). This makes the theory more difficult but ensures at the end the spatial separation of the twin photons. Let us take the most common non-linear crystal used for SPDC up to now: barium borate $\beta - Ba(BO_2)_2$ or BBO. It is a negative (meaning that the direction having a low refractive index is the fast axis. At right angles to it is the slow axis, with a high index of refraction, $n_e < n_o$) uniaxial crystal (meaning it has an optical axis OA) with two different indices $n_e$ and $n_o$ for extraordinary and ordinary indices. BBO is transparent from 190 nm to 3300 nm and with indices given by the Sellmeier equations which are, for example, for type I down-conversion ($e \mapsto o + o$):

$$n_o = 2.27405 + \frac{0.0184}{\lambda^2 - 0.0179} - 0.0155\lambda^2 \tag{33}$$

$$n_e = 2.3730 + \frac{0.0128}{\lambda^2 - 0.0156} - 0.0044\lambda^2 \tag{34}$$

with $\lambda$ given in $\mu m$. We recall the relation of the extraordinary index of refraction with the angle $\theta$ from the optical axis is given by the ellipsoid equation for indices:

$$n^2(\theta) = \frac{n_e^2 n_o^2}{n_e^2 \cos^2\theta + n_o^2 \sin^2\theta} \tag{35}$$

So typically, when the pump beam enters the crystal with a certain angle $\theta$ with the optical axis OA, the extraordinary index will be different for different angles, as opposed to the ordinary index. For BBO, phase-matching conditions occur with an extraordinary beam only and one can have two types of phase-matching:

$$e \mapsto o + o \quad \text{type I}$$

$$e \mapsto e + o \quad \text{type II}$$

$$e \mapsto o + e \quad \text{type II}$$

Figure 1-d represents a type II configuration. Figure 2 represents the two curves for the Sellmeier equations of BBO with $n(\theta)$ for $\theta = 90°$ (in black) and $n_o$ (in red) as a function of the wavelength $\lambda$. We will take the simple case where we have collinear phase matching and with a pump at 400 nm and hope to get a pair of 800 nm photons. The pink curve is $n(\theta = 31.7°)$ where we can see that for a pump beam at 400 nm (very common in the laboratory) we can have phase matching at 800 nm (the two grey circles) where we fulfill a condition where $n_e(\omega_s) = n_o(2\omega_p)$. This is the phase matching condition for this particular case at these particular wavelengths for type I and in a collinear configuration. This simple case is for illustration but the most general case cannot be evaluated without the help of numerics. The idea is to solve equation (5) in a vector form $|\Delta \vec{k} = \vec{k}_p - \vec{k}_s - \vec{k}_i| = 0$. In order to do so, one needs to work in spherical coordinates and determine 9 parameters to fulfill the phase matching conditions given by the equations (4) and (5): 3 parameters for the wavelength $i, s, p$, 3 angles $\theta$ for each wave and 3 angles $\varphi$ for each wave. Clearly one can set $\lambda_p, \lambda_s, \lambda_i$ for the 3 waves if we want to work at certain wavelengths and in the degenerate case. We can also decide to set $\theta_s$ for instance (and not be in the collinear case such as in Fig. 1-d) and we can show that for a uniaxial crystal,

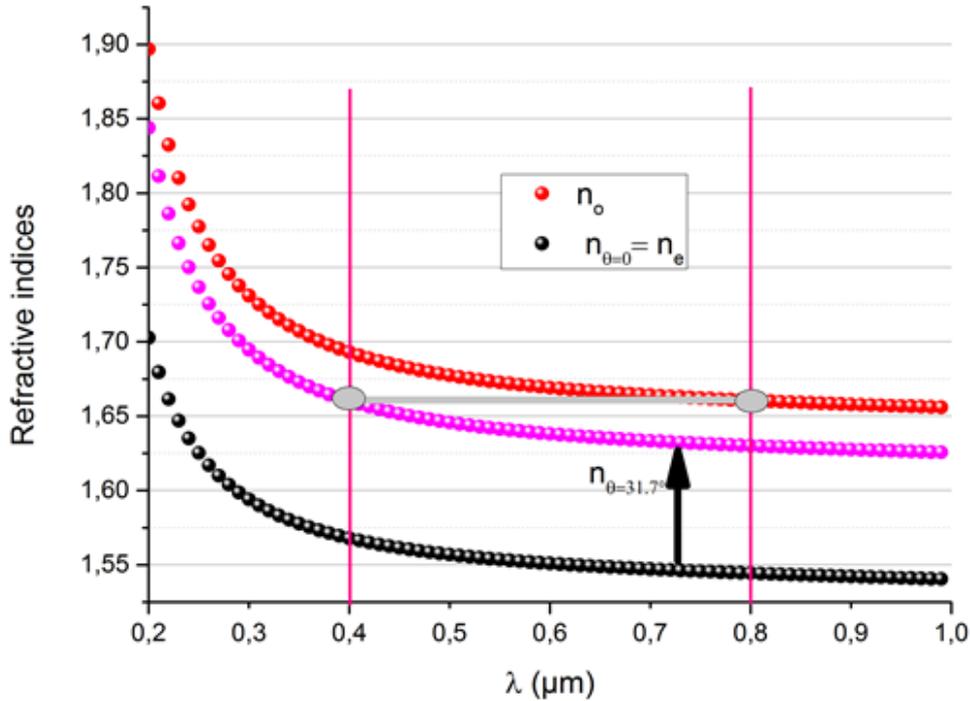

**Figure 2:** Graphs representing the Sellmeier equations of BBO with $n_e(\theta)$ for $\theta=90°$ (in black) and $n_o$ (in red) as a function of the wavelength $\lambda$. The pink curve is $n_e(\theta=31.7°)$. We have phase-matching between 400 nm and 800 nm (the two grey circles) when we fulfill a condition where $n_e(\omega_s)=n_o(2\omega_p)$. This is the phase matching for SHG condition for this particular case at these particular wavelengths for a type I non-linear BBO crystal and in a collinear configuration.

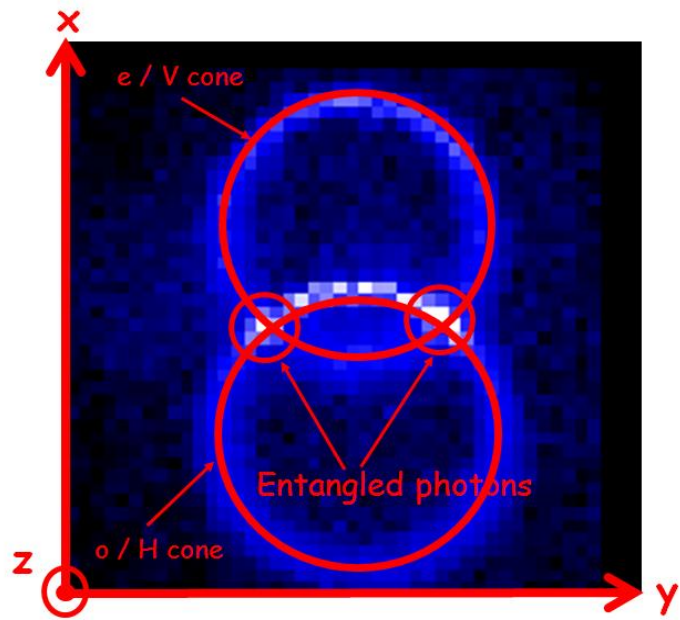

**Figure 3:** Graph representing the two outcoming cones of SPDC for a type II BBO crystal. This is an experimental transverse cut of SPDC at 810 nm behind a BBO crystal. The two circles represent the two extraordinary and ordinary beams. In this particular case, we can observe two intersection points (where polarization entanglement occurs). We can also observe that one circle is slightly bigger than the other one due to crystal tilting and meaning that the two wavelengths for signal and idler are slightly different.

there is no dependence with $\varphi_p$ and that there is the relation $\varphi_s = \pi + \varphi_i$. We are finally left with 3 parameters $(\theta_p, \theta_i, \varphi_s)$ to determine using equations (4) and (5). For more information in the calculation part, see the article from NIST by Boeuf *et al.* from 2000 [22]. The following website maintained by NIST: spdcalc.org allows you to enter your own parameters and gives you all sorts of information such as the phase-matching curves or the joint spectral distribution. In particular, one derives from these calculations that for a type II SPDC in BBO with a pump laser at 405 nm, giving rise to 810 nm down-converted photons, we have an angle of about 3° between the idler and the signal waves coming out of the crystal in a cone shape (see Figure 1-d for a schematic). Figure 3 presents an experimental result of the double cone emission of SPDC in type II BBO. In this case, we have two different circles as opposed to the type I case. Figure 3 represents the notorious picture of the parametric down-converted cones coming out of the crystal with photons with orthogonal polarisations for the two different cones (extraordinary e/V cone and ordinary o/H photons). This is an experimental transverse cut of SPDC at 810 nm behind a BBO crystal done by scanning across a 2D plane parallel and away from the nonlinear crystal using an optical fibre linked to a single photon detector. We will see later on that it can lead to polarisation entangled pairs of photons. For that, one will need to take two things into account that are inherent to SPDC: the temporal walk-off and the spatial walk-off. For the temporal walk-off, also known as the longitudinal walk-off, it simply shows that as the crystal has two indices along which the 'e' and the 'o' photons follow, they will travel differently (different group velocities) and be temporally different when coming out of the crystal (the length of the crystal is typically on the order of 1 mm). One photon will come out of the crystal before the other one. This temporal difference is given by:

$$\Delta t = \left| L_c \left( \frac{1}{n_e} - \frac{1}{n_o} \right) \right| \tag{36}$$

The spatial walk-off, or transverse walk-off is due to the fact again that the material is birefringent and as such the energy propagation or the Poynting vector is not necessarily in the same direction as the wavevector of the down-converted photons. This is the case for the ordinary wave o but not for the extraordinary wave e and we thus end up with an angle $\rho$ between the two waves at the output of the crystal given by:

$$\rho = -\frac{1}{n_e}\frac{dn_e}{d\theta} \tag{37}$$

Once again, this will be very important for the production of entangled photons as one must not have any kind of information (temporal or otherwise) in order to produce pure entanglement. There are easy ways to avoid these 'labelling' by using compensating crystals (see [8] for instance). One more thing we should emphasize is the fact that for spontaneous parametric down-conversion, the conservation of momentum has an uncertainty to it (equation (5)) and we have rather:

$$\vec{k}_3 = \vec{k}_2 + \vec{k}_1 + \Delta\vec{k} \tag{38}$$

and the equation still holds. That is the reason why, these two cones actually form two rings when one makes a transversal view-cut along the propagation direction of the SPDC photons (see Fig. 3). One way to overcome this situation is to select spectrally with narrow-band filters (from 1 to 10 nm) the down-converted photon pairs, on top of a spatial (with angles) and temporal (with coincidence detections) selection [11,23].

## IV- SPDC used for applications

*Single-photon source*

The first striking effect with SPDC is the fact that we end up with twin photons created at the same time and which are indistinguishable. One foresees that this could be useful for some low-signal level communications and was predicted by Mandel and Rarity (c. 1984) as a joint measurement can ensure their 'twin-ness'. Derive from that, the first application was the localisation of a one-photon state using SPDC by [24]. This led to heralded single photon sources using SPDC providing strong photon antibunching which is again a clear quantum property of light. This is good for quantum information processing (QIP) although more for proofs of principle. Real single photon sources such as quantum dots, NV centers, atoms, etc... are still preferable. Figure 4-a is an example of an heralded single photon source where the authors used this time collinear source of orthogonal twin photons that were separated in polarisation [12]. One photon was detected (idler photon in Fig 4-a) and served to herald the arrival of the signal photon on the other side which was then detected using a standard Hanbury-Brown and Twiss (HBT) interferometer known in quantum optics for the determination of single photons. Fig 4-b presents the photon antibunching characteristics of this system as a function of the time coincidence window for the detection. We can show that the quality of this autocorrelation $g^{(2)}$ degrades as the intensity of the pump increases and that is due to the fact that the emission of spontaneous parametric down-conversion follows a Poisson statistics as discussed previously with equation (24). Thus SPDC has a limited use as a source of single photons as multiple pair creation is always present.

*Source of indistinguishable photons*

In 1987, the group of Leonard Mandel realised a two-photon interferometry and observed this time a photon bunching effect. This experiment consisted in creating two twin photons by SPDC and then made them interfere on a beamsplitter (cf Figure 5). Quantum-mechanically, one can show that the two outcomes where the photons come out in different ports interfere destructively and remain only the cases where the photons come out together from the BS, they bunch. This shows bosonic properties for photons due to the fact that the quantum operators of annihilation and creation of photons are used (oppositely, fermions like electrons would go away from each other). The photons are said indistinguishable as they must be from the same optical mode in order for the interferences to occur and this is called the Hong-Ou-Mandel effect [10].

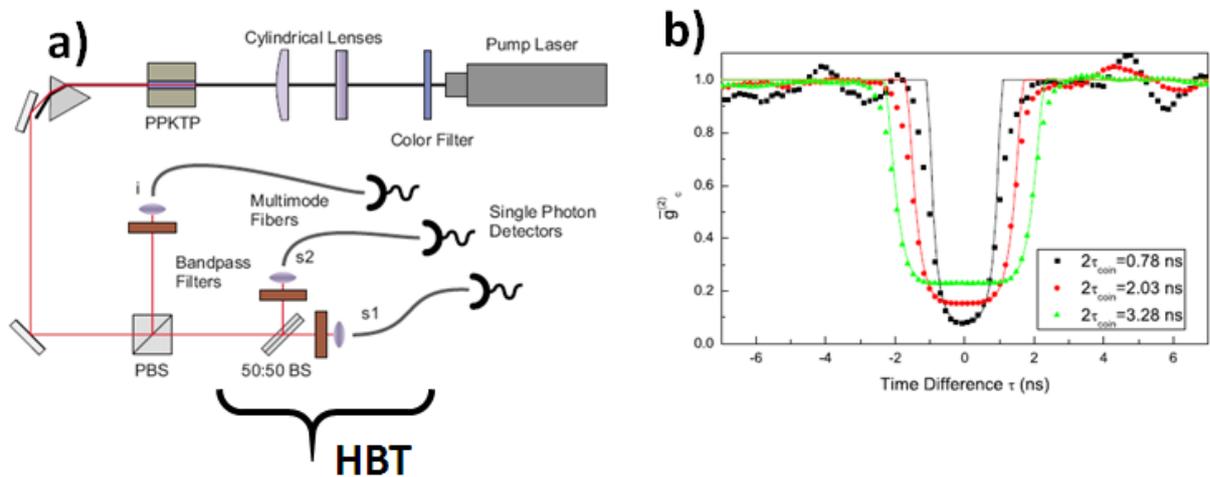

**Figure 4: a) Example of an heralded single photon source using SPDC from a periodically poled KTP crystal (PPKTP). The crystal produces a pair of collinear photons with orthogonal polarisations which are first separated by a polarizing beamsplitter (PBS), so that the idler photon serves as an heralding event for the signal photon sent onto a non-polarising beamsplitter (BS) for a measurement of the $g_c^{(2)}$ function (so-called HBT). b) $g_c^{(2)}(\tau)$ function for 3 different coincidence windows showing $g_c^{(2)}(\tau) \approx$ 0.07 for the lowest coincidence window [12].**

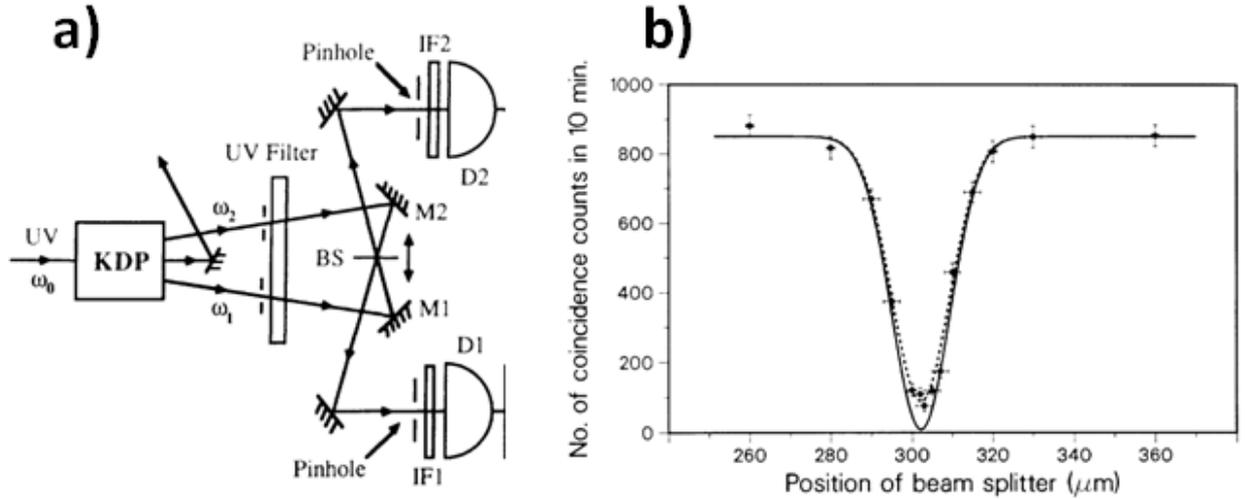

**Figure 5: a)** Experimental set-up of the first two indistinguishable photon experiment pioneered by Hong, Ou and Mandel [10]. This was done by doing SPDC in a KDP in a non-collinear phase-matching condition. The BS is moved to change the delay between the arrival times of the photons onto the BS. **b)** Coincidence measurements of the photon pairs detected by D1 and D2 as a function of the path delay at the BS with a clear dip observed.

*Entangled photons*
Following these experiments, researchers used parametric down-converted photons for creating entangled photons in polarization for violating Bell's inequalities. Figure 6 presents the original article where they managed, using a type-II BBO crystal, to create two intersections of the output cones by tilting their BBO crystal (see also Figure 3). They then selected the intersection of the two parts of the cones of the twin photons with orthogonal polarisations. As with the right conditions it is impossible to tell which photon is which anymore, you cannot predict the result of the measured polarisation of the photon at one intersection. What you do know though is the fact that parallel measurements will give you perfect anti-correlation in polarisation. Figure 1-d is a schematic of the two output cones from idler and signal and Figure 3 provides experimental data where we have a transverse cut of the two twin beams produced. At the intersection, one ends up with the following entangled state:

$$|\psi\rangle = \tfrac{1}{\sqrt{2}}\left(|H_i\rangle|V_s\rangle + e^{i\phi}|V_i\rangle|H_s\rangle\right) \tag{39}$$

where *H* is for horizontal polarization and *V* is for vertical polarisation, *s* and *i* stand for signal and idler like before and ϕ is a phase that can be modified. Figure 6-a represents a schematic of the experiment carried on by Kwiat *et al.* [11], while Fig 6-b shows a diagram of the geometry of the process. Figure 6-c demonstrates clearly that there are strong correlations in polarization between the two entangled photons. This entangled state and this technique was used to violate Bell's inequalities and brought up a lot of discussions regarding the so-called Einstein-Podolsky-Rosen (EPR) paradox.

*Quantum teleportation*

One of the most famous use of entanglement is certainly when it was applied to demonstrate quantum teleportation. The idea of teleportation is that two protagonists (usually called Alice and Bob) share a common pair of EPR particles (in this case photons) that they will use in order to teleport a quantum state (which does not have to be known) from Alice to Bob. Figure 7-a shows the general principle of creating a pair of entangled/EPR photons (2 and 3) distributed between Alice and Bob and Alice uses her photon 2 to make a so-called Bell state measurement (BSM) with the photon to be teleported (photon 1) [13]. Using a classical channel of information, Alice tells Bob what BSM-type of measurement she made (but not the result) so that Bob can act accordingly on his photon n°3 and recover the quantum state of the teleported photon 1. Figure 7-b represents the experimental set-up developed in 1997 at the University of Innsbruck. A first pair of entangled photon is created by a pump beam where photon 3 will be going to Bob and photon 2 will be going to Alice. Alice then makes a BSM on a beamsplitter with photon 1 to be teleported which is in fact coming from the creation of another pair of EPR photons by sending back the pump beam into the crystal. This time one photon of the pair is photon 1 and the other one will be used to herald the arrival of photon 1 for Alice to make the BSM. We note that in order to synchronise the whole process, the authors use a pulse excitation with a well-known arrival time for each time (within the pump pulse) and photon 4 is used to herald the arrival of photon 1 to be teleported. Finally, the equivalent classical communication between Alice and Bob is done by the fact that they have to communicate together to see when they made a coincidence measurement. Using a similar set-up, entanglement swapping was also realized [25] and now people try to store single photons and entangled photons into quantum memories from specific materials [26].

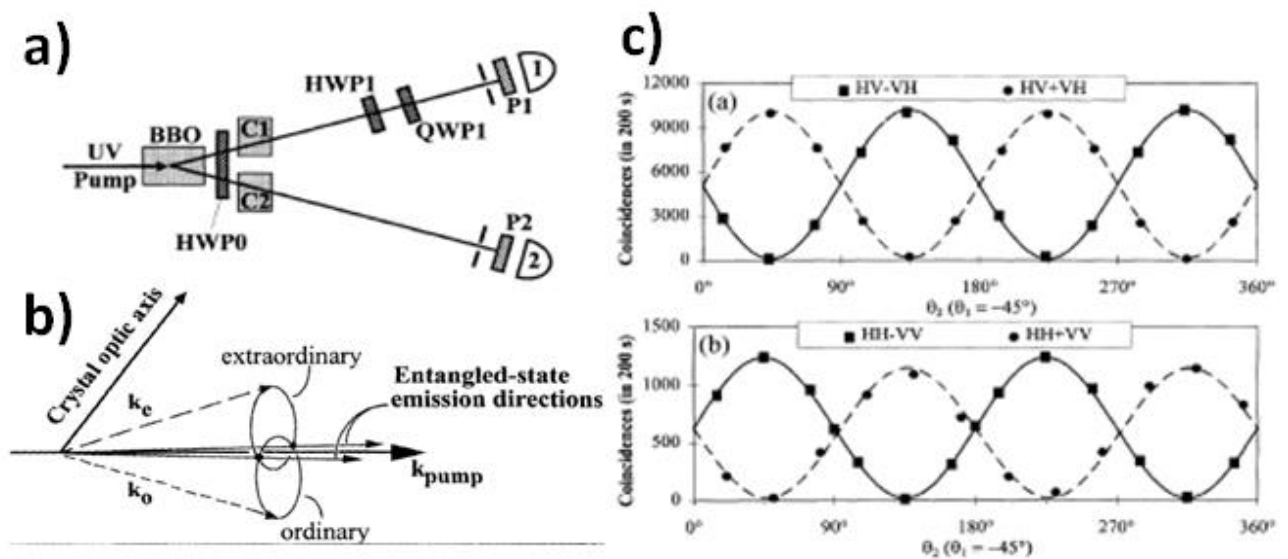

**Figure 6: a) Set-up of the first experiment of two entangled photon experiment done by Kwiat et al. where a BBO crystal is pumped at 351 nm for SPDC in a non-collinear critical phase-matching condition. b) Schematic of the famous two SPDC cones coming off the crystal with orthogonal polarisations and spatial selection to select only the photons at the intersection. c) Visibility measurements leading to a Bell parameter S > 2.5 and proving that the photons are indeed entangled [11].**

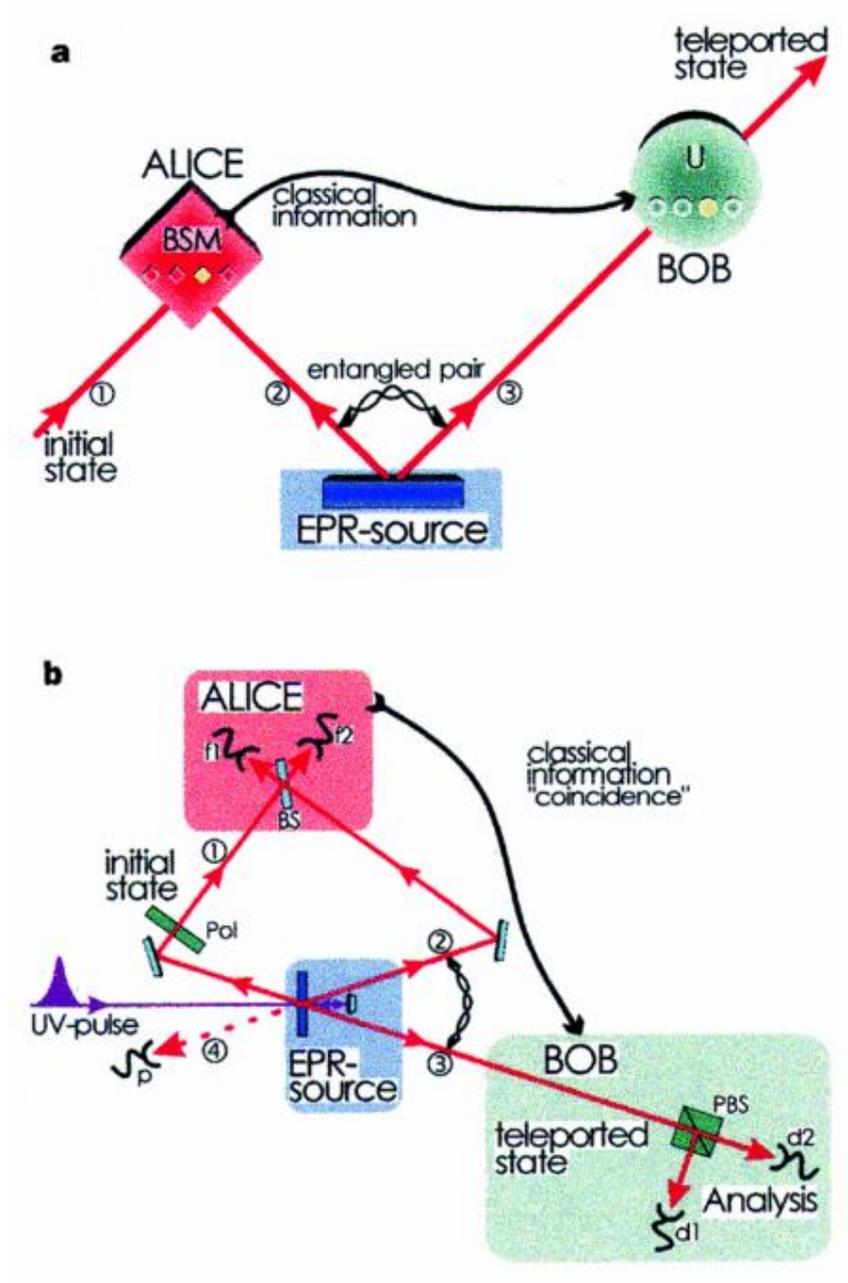

**Figure 7: a)** Principle of the teleportation experiment where we send an initial state 1 and Alice proceed in doing a Bell-state measurement (BSM) with one entangled photon from the EPR source (2). This initial state is the one to be teleported. The other entangled photon of the EPR pair (3) is sent to Bob who receives a classical information from Alice onto what measurement he should do in order to recover Alice's initial state, thus effectively teleporting the initial state 1 to state 3 from Alice to Bob. **b)** Schematic of the experiment where a strong pulse of pump photons first creates an EPR entangled pair (2 and 3) to share between Alice and Bob. This pump pulse of light is then sent back again into the non-linear crystal to create another pair of SPDC photon (1 and 4) where one photon of the pair will be the initial state to teleport (path 1) and the other one will be used as a trigger to create the heralded photon source (4) [13].

*Quantum cryptography*

This teleportation experiment and some others at the time certainly sparked a lot of interest in the broad field of quantum information. In particular, it was shown earlier on that using entanglement and the no-cloning theorem in quantum mechanics, one could make a very secure protocol based on quantum cryptography (at least on paper). The basic idea proposed is that the polarisation state of a photon (say horizontal or vertical) can be thought of as a quantum bit of information, or a qubit. Thus a pair of EPR photons can represent a two-qubit source. Figure 8-a and –b presents two experiments realising quantum cryptography using entangled photons. In Figure 8-a, the authors use polarisation entangled photons [14] but in Figure 8-b, the group from Nicolas Gisin in Geneva used the so-called time-bin entanglement based on James-Franson's idea from 1991 [27]. In this case, the idea is simply to create twin photons (say A and B) and to send each of them in a Mach-Zehnder interferometer with a long (L) and a short path (S). Then if one gets a coincidence measurement, assuming that beforehand we use a pulse pump beam itself in a superposition of L and S paths, there is no way to tell whether the photons took both the short paths or the long paths [28]. All we know is that they have both taken the same path in order to get a coincidence measurement. This very powerful method of entangling photons allowed the Geneva group (and others thereafter) to use non-degenerate SPDC to get one photon at the pair at, say 900 nm in the silicon detection window, and a photon at 1550 nm [29]. This wavelength is not chosen randomly as it corresponds to the wavelength used in the world communication network thus bringing quantum cryptography into the scene of a mature application.Of course, science is not that easy and other problems have been encountered like inefficient detectors, problem with maintaining stable interferometers etc… The carrier of information has been tested on optical tables but also in fibers (visible or telecom) and in free-space. Now there are projects to point to satellites to do satellite communication quantum cryptography [30].

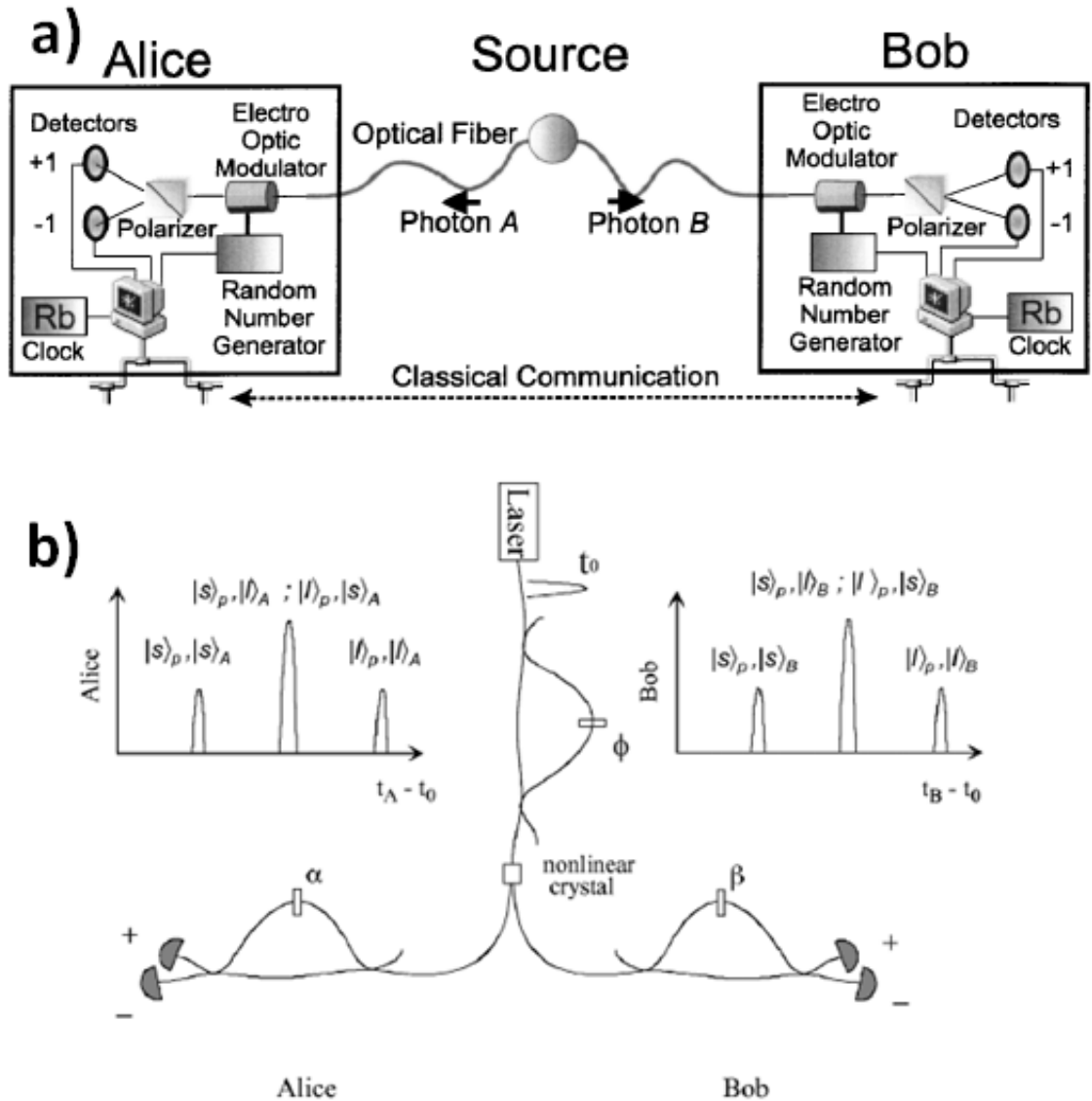

**Figure 8: a) Schematic of the quantum cryptography protocol using polarisation entangled photons using the SPDC method of Kwiat *et al.* (Figure 6). Each photon from a pair is sent to Alice and Bob where they verify that the photons are indeed entangled in order to use them as an encryption key. H polarization can be the "+1" bit of information and V polarization can be the "-1" bit of information [14]. b) Quantum cryptography was also realised in telecommunication fibres where polarization is not preserved. For that, Tittel *et al.* used the so-called time-energy entanglement or time-bin entanglement to generate entangled photons [28].**

*Quantum computation*

Following these remarkable advances in quantum communication, it did not take long before SPDC was used for running experiments on quantum computation. This is still a very much on-going task where groups around the world demonstrated basic algorithms, error-correction codes and more using entangled photons from SPDC. SPDC can only generate two quantum bits (qubits) at the time (at least in polarisation) and people have been working actively in order to get 2, 3 up to 10 pairs of entangled photons in a given experiment [11,31,32,33,34,35,36]. Figure 9 plots the number of polarisation entangled photons created by SPDC as a function of time. At this linear rate, we will reach 20 entangled photons in 2037. Unfortunately, the trend does not follow Moore's law and it seems pretty clear it may even reach a plateau at some point using this technique. Nevertheless, it is still primordial to show basic quantum information protocols using SPDC and entangled photons. Also shows the need for new sources of entangled photons. The first real practical implementation of an optical gate was realized in 2003 with entangled photons [15]. This was the first experimental all-optical quantum controlled-NOT gate (CNOT gate). This experiment and many others followed the so-called KLM proposal (published by E. Knill, R. Laflamme and G. Milburn in 2001 and it was the first article published in Nature of the twenty first century [37]. The KLM proposal showed that using non-deterministic gates, single photons, efficient and discriminating single-photon detectors, it was possible to perform an all-optical computer using only linear optics such as beamsplitters, wave-plates etc... Figure 10-a presents the basic principle of this CNOT gate with a control qubit and a target qubit. Each qubit can have a 0 or a 1 state and the CNOT gate tells you if the control qubit is in the 0 state, then the state of the target state is unchanged whereas if the control qubit is in the 1 state, then the target state will have its state flip from 0 to 1 or from 1 to 0. This scheme used by O'Brien *et al.* was a coincidence basis gate which is non-deterministic [15]. Fig 10-b represents the so-called dual-rail encoding where spatial superposition with a beamsplitter is realised in order to implement this scheme. This, combined with entangled photons in polarization and conditional detections showed that the gate was realized $1/9^{th}$ of the time. Since then, this scheme was improved. Using SPDC, it has been used to implement basic concepts in quantum processing through the proof of CNOT gates [15], Shor's algorithm [38], one-way quantum computation [39] or boson sampling type of gates [40]. There is a wide literature on the subject and it is still very much on-going [4].

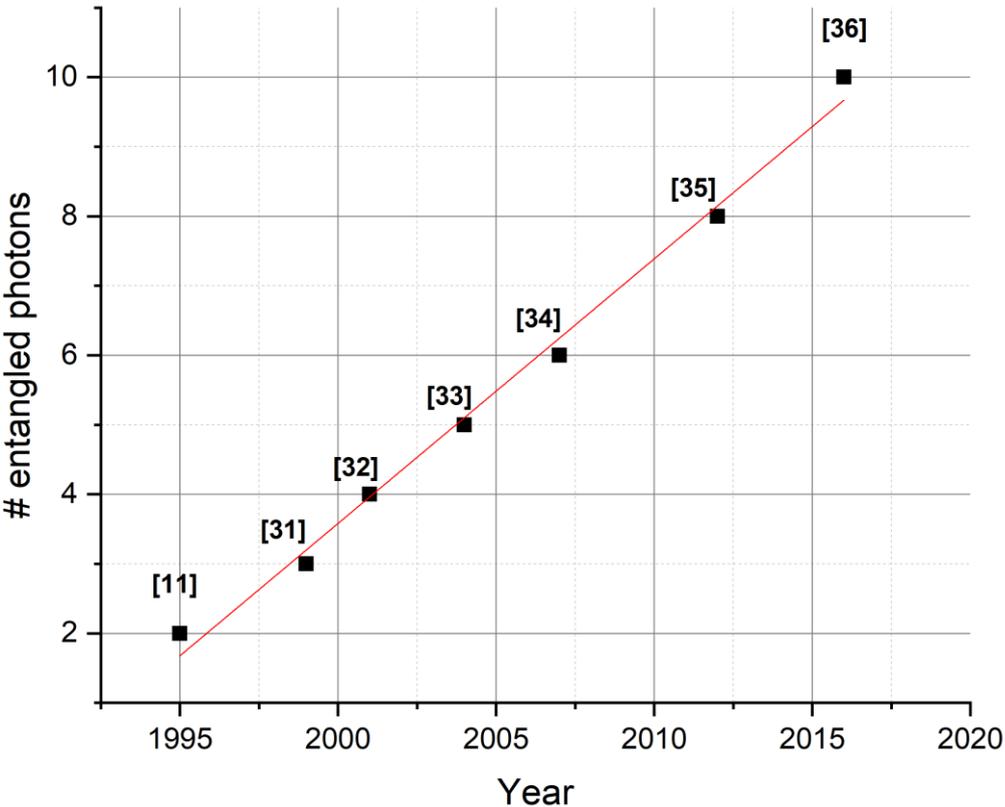

**Figure 9: Number of polarisation entangled photons in time using SPDC. The black squares are experimental results with the reference and the red line is linear fit. At this linear rate, we will reach 20 entangled photons in 2037.**

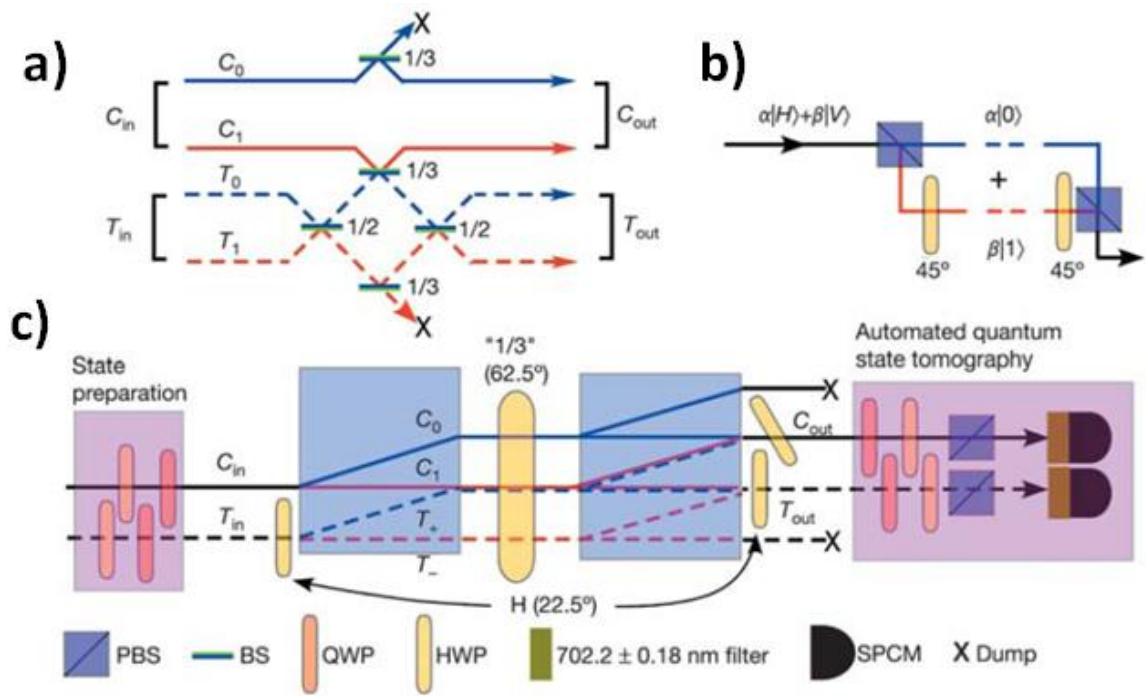

**Figure 10:** a) Schematic of the CNOT gate realized in [15]. b) Polarisation-encoded photonic qubits can be converted into spatially encoded qubits, suitable for the gate shown in a) and c) schematic of the experimental CNOT gate. Pairs of energy degenerate photons are incident from the left of the diagram. They then follow the gate procedure in the middle before being detected on the right-hand side [15].

*Quantum metrology and more*
SPDC allows also other types of less-obvious applications. One of them is the use of entanglement in order to perform super-resolution phase measurements [17]. The basic idea is to create a so-called NOON state written as $|N,0\rangle + |0,N\rangle$ where N is the number of photons in a given mode. In this case, using interferometry, one can show a phase resolution given by $\Delta\varphi = 1/\sqrt{N}$. The concept of 'quantum lithography' was thus born although there is still much more improvement to be done before this concept can actually be put in practice. One example was to use entangled photons for measuring protein concentration [41].

Another idea that came out fairly earlier on when spontaneous parametric down-conversion photons were characterised, is the idea that one could use twin photons for metrology. The concept of quantum metrology uses light for setting the standards that define units of measurement (the candela) for light. The quantum candela project aims at developing standards for photon metrology from the signal level of existing radiometric standards and reach the smallest grains of light which are single photons. Through photon correlation measurements, it was shown that one could calibrate very accurately photons detectors, again a very important application for radiometric metrology [16,5,42].

Finally, we must mention that originally, parametric fluorescence and entangled photons generated with them where used for testing the very foundations of quantum mechanics such as the wave-particle duality, non-locality via the violation of Bell's inequalities where in 2015, 3 experiments finally closed all the loop-holes associated to these tests and put an end to hidden variables introduced by John Bell in the 60s. Two of them used SPDC for that purpose [43,44]. Born's rule of quantum mechanics was also tested with an heralded single-photon source using SPDC going through a three slit interferometer [6]. This was done in order to show that quantum interference is a two-amplitude effect and that no three-amplitude term exists in quantum mechanics.

## V- Future for SPDC

The future for SPDC is still very bright and nowadays, SPDC is used as a tool more than anything, for various applications and we can classify the applications into three main categories: a-foundations of physics and in particular of quantum mechanics, b-quantum information processing and c-quantum metrology via the quantum candela project and quantum sensors. Within QIP, we can split into 3 sub-categories: quantum cryptography, quantum computing and quantum communications. There are other secondary potential applications for optical lithography but the odds that any application will come out of this are pretty low, quantum imaging, also very narrow field and quantum sensing which has a higher potential.

Now the challenge of tomorrow is to actually engineer new technologies to create efficient SPDC and not just using the natural crystal birefringence. Several groups are attempting to do so by several means, either by using confined light into optical fibres [45], in SOI waveguides [46] but also by careful engineering of semiconductor Bragg-type of structure in order to obtain parametric down-conversion directly on an optical chip [47]. By the same token, using two four-wave mixing processes, Silverstone *et al.* managed to create SPDC photons directly on a chip made of silicon-on-insulator [46]. A direct 3-photon has also been shown using cascaded SPDC [48]. Recently, an electrically injected photon-pair at room temperature was even demonstrated with is the first evidence of SPDC obtained by electrical pumping rather than optical pumping [49]. More recently, we should mention the very interesting work on enhancing SPDC effect using non-linearities in metamaterials [50]. As such, it is safe to say that research in SPDC has thus a bright future ahead.

ACKNOWLEDGEMENTS

The author would like to point out that most of this work was done when the author was at the University of Oxford from 2000 to 2002 and as such, would like to thank, Dik Bouwmeester, John Rarity, Christoph Simon, Antia Lamas-Linares, Gabriel Durkin and John Howell but also Gregor Weihs, Chris Erven, Rolf Horn and Kevin Resch for fruitful and sometimes very entangled discussions. The author thanks L. Le Cunff for Figure 1-d and C. Altuzarra and S. Vezzoli for working on the data of Figure 3.